\documentclass[useAMS,usenatbib]{mn2e}
\usepackage{natbib}
\usepackage{graphicx}
\usepackage{subfigure}
\usepackage{amssymb,amsmath}
\usepackage{symb}    

\title[Most massive halos with Gumbel statistics]{Most massive halos with Gumbel statistics}
\author[O. Davis, J. Devriendt, S. Colombi, J. Silk \& C. Pichon]{O. Davis$^1$\thanks{E-mail:
olaf.davis@astro.ox.ac.uk}, J. Devriendt$^1$, S. Colombi$^2$,
J. Silk$^1$ \& C. Pichon$^{1,2}$\\
$^{1}$Astrophysics, University of Oxford, Denys Wilkinson Building, Keble Road, Oxford OX1 3RH, UK\\
$^{2}$Institut d'Astrophysique de Paris, CNRS UMR 7095 \& UPMC, 98bis bd Arago, F-75014 Paris, France}
\begin{document}

\date{}

\pagerange{\pageref{firstpage}--\pageref{lastpage}} \pubyear{2010}

\maketitle

\label{firstpage}

\begin{abstract}
We present an analytical calculation of the extreme value statistics for dark matter
halos - that is, the probability distribution of the most massive halo within some region
of the universe of specified shape and size. Our calculation makes use of the
counts-in-cells formalism for the correlation functions, and the halo bias derived
from the Sheth-Tormen mass function.

We demonstrate the power of the method on spherical regions, comparing the results to
measurements in a large cosmological dark matter simulation and achieving good agreement.
Particularly good fits are obtained for the most likely value of the maximum mass
and for the high-mass tail of the distribution, relevant in constraining cosmologies
by observations of most massive clusters.
\end{abstract}


\section{Introduction}

Extreme value (or Gumbel) statistics are concerned with 
the extrema of samples drawn from random distributions. 
If a large number of samples are drawn from some distribution,
the Central Limit Theorem states that their respective means will
follow a distribution which tends, in the limit of large sample size,
to a member of the family of normal distributions.
Analogously, the maximum (or minimum) values $u$ of the samples will have a distribution
whose large sample size limit -- where such a stable limit exists \footnote{Although it can be shown
that where a stable limiting distribution exists it will take the form (\ref{eq:GEV}),
certain pathological distributions give no such limit. For our purposes it is sufficient
to note that the limit indeed exists for distributions which are of exponential type, meaning
the cumulative distribution function $F$ obeys $\lim_{x\rightarrow\infty} d/dx \{(1-F(x))/F'(x)\} = 0$, and that this class
includes all physical distributions relevent to our applications.}
-- is a member of the family of Generalised Extreme Value (GEV) distributions
as detailed by \citet{Gumbel}:

\begin{equation}
-\ln P_{\rm GEV}(y) = (1+\gamma y)^{-1/\gamma}, \quad y=(u-\alpha)/\beta.
\label{eq:GEV}
\end{equation}
The shape parameter $\gamma$ is sensitive to the underlying distribution from
which the maxima are drawn, while $\alpha$ and $\beta$ are position
and scale parameters. 

Despite their wide use in other fields, extreme value statistics have historically seen
very little application in astrophysics; some exceptions are the work of \citet{BhavsarBarrow}
on the brightest galaxies in clusters and the study of \citet{ColesP}
on the hottest hot spots of the cosmic microwave background temperature fluctuations.

The past year or so, however, has shown a resurgence of interest in the application of extreme
value statistics to cosmology and questions of extreme structures, as revealed either in
the clustering of galaxies \citep{Antal&al,Yaryura&al}, the prevalence of massive clusters
\citep{HolzPerlmutter,Cayon&al10} or the temperature extrema of the CMB
\citep{Mikelsons&al}.

In this paper, we are interested in the dark matter halos of massive
galaxy clusters. The number density of extremely massive clusters is indeed 
a sensitive probe of the effects of the underlying cosmological model
and laws of physics on large scales \citep{Mantz&al2}. These include for instance  
the equation of state of dark energy \citep{Mantz&al1}, the possibility of modified
gravity \citep{Rapetti&al} the physical properties of neutrinos \citep{Mantz&al3} and
primordial non-Gaussian density fluctuations 
\citep{Cayon&al10}. Although the majority of the previously mentioned studies
focus on the growth of massive clusters, simply
knowing  the mass of the most massive cluster in a survey can already provide
strong constraints on cosmology \citep{HolzPerlmutter}.

In the present work, we outline an analytical derivation of the extremal
halo mass distribution in standard cosmologies with Gaussian initial 
conditions. Rather than taking a phenomenological approach, we aim to
predict the distribution of the most massive halo in a region for
any specified combination of power spectrum, cosmological parameters and region size and
shape.  The paper is organized as follows. In  \S~\ref{sec:theory} we outline
the basics of our method for obtaining an analytical expression of the
Gumbel distribution of most massive clusters masses and make an
explicit link with eq.~(\ref{eq:GEV}). 
In \S~\ref{sec:HOR}, the theoretical predictions are
checked against measurements in a very large $N$-body
cosmological simulation. Finally, \S~\ref{sec:conclusion} follows with
a short summary of the main results and conclusions. 

\section{Theory}
\label{sec:theory}
Consider a large patch of the universe, which can be thought of as representing
the space covered by a volume-limited sample of clusters, and denote by $m_{\rm max}$
the mass of the most massive dark matter halo in that patch.
We wish to study analytically 
the Gumbel statistics, that is the probability distribution function
$p_{\rm G}(m_{\rm max}){\rm d}m_{\rm max}$ of the values taken by $m_{\rm max}$ if we
sample a large number of such patches. Obviously, this distribution will depend on
the size and shape of the patch, as well as its redshift.

\subsection{General expression of the Gumbel statistics}
\label{sec:gumbel}
Let us define the cumulative Gumbel distribution by
\begin{equation}
P_{\rm G}(m)\equiv {\rm Prob.}(m_{\rm max} \le m)
\equiv \int_0^{m}p_{\rm G}(m_{\rm max}) {\rm d}m_{\rm max}.
\end{equation}
Such a probability is also
the probability $P_0(m)$ that the patch is empty of halos of mass above
the threshold $m$ \citep{Colombi}, hence
\begin{equation}
p_{\rm G}(m)=\frac{{\rm d}P_0}{{\rm d}m}.
\end{equation}
Note that this assumes that there are no significant edge
effects, i.e. that the boundaries of the catalog do not cross
too many clusters.  This effect is negligible if the patch is
large compared to the halo size (and sufficiently compact). 

If halos are unclustered then the void probability follows simply
from Poisson statistics,
\begin{equation}
P_0(m)=\exp(-n V),
\label{eq:p0poisson}
\end{equation}
where $V$ is the volume of the patch and \mbox{$n=n( > m)$} the
mean density of halos above mass $m$, with the appropriate
spatial average with redshift made implicit \mbox{$n(>m) = \langle
n[>m,z(x)] \rangle_{x \in V}$}.

We expect the Poisson limit to be reached for patchs of size 
above a few hundred Mpc, where the matter distribution
becomes homogeneous. Below this patch size, however, halos are significantly
clustered.  In that case, the calculation of the void probability can be performed using
a standard count-in-cell formalism if the connected $N$-point
correlations functions, $\xi_N^{\rm h}(x_1,\ldots,x_N)$, of halos
above the threshold are known \citep[e.g.,][]{SzapudiSzalay,BalianSchaeffer89}.
The superscript h in the previous expression indicates halo
correlation functions, while the naked $\xi_N$ refer to correlations 
of the underlying matter density field. 
Since deviations from Poisson behavior occur only for moderate
patch sizes, the complex lightcone effects on the  correlations
induced by the evolution of clustering with redshift inside the patch \citep[e.g.][]{Matsubara} 
can safely be neglected. 

In particular, one can define the averaged correlations over a patch of
volume $V$:
\begin{equation}
{\bar \xi}_N^{\rm h} \equiv \frac{1}{V^N} \int_{V} {\rm d}^3x_1 \cdots
{\rm d}^3 x_N
\xi_N^{\rm h}(x_1,\cdots,x_N),
\label{eq:xin}
\end{equation}
and the typical number of halos above the threshold $m$ 
in excess to the average in overdense patches as:
\begin{equation}
N_{\rm c} \equiv n V {\bar \xi}_2^{\rm h}.
\end{equation}
In the high-$m$ limit,
the void probability can be written
\begin{equation}
P_0(m)=\exp\left[-n V \sigma(N_{\rm c}) \right],
\label{eq:p0clustered}
\end{equation}
where the function $\sigma(y)$ reads
\begin{equation}
\sigma(y)=\left( 1 + \frac{1}{2} \theta \right) {\rm
  e}^{-\theta},\quad \theta {\rm e}^{\theta}=y,
\label{eq:sigma}
\end{equation}
\citep{BernardeauSchaeffer}.
Note that, as pointed out by these authors, this expression for $\sigma(y)$ 
follows from a specific hierarchical behavior of higher-order correlation
functions of very massive halos at large separations, ${\bar
  \xi}_N^{\rm h} \simeq N^{N-2} ({\bar \xi}^{\rm h}_2)^{N-1}$.

We now proceed to specify the cumulative halo number density and
the average two-point correlation function of halos in order to fully determine the Gumbel statistics.

\subsection{Halo number density}
\label{sec:no_density}
The number density $n(m,z)$ of halos at a given mass $m$ and
redshift $z$, a.k.a. the halo mass function,  we adopt is the one calculated by \citet{ST99}. It is based on 
a modification of the original model of \citet{PS74}, which links the statistics of the initial matter density
field to the distribution of virialized dark matter halos through a
spherical top hat description of their gravitational collapse. As a
result, this mass function can be expressed as a universal function of $\nu \equiv (\delta_{\rm c}/\sigma(m,z))^2$,
where $\sigma(m,z)^2$ is the variance of the initial density field smoothed over
spheres of radius $R(m)$ containing an average average mass $m$ linearly extrapolated to
$z$, and $\delta_{\rm c}$ is the critical overdensity threshold needed
to turn an initial spherical top hat density perturbation into a collapsed halo at redshift $z$. 
More specifically, the number density of mass-$m$ halos is given by:
\begin{equation}
m^2\frac{n(m,z)}{\bar{\rho}_{\rm m}}\frac{{\rm d}\log m}{{\rm d}\log\nu} =
A\left(1+(a\nu)^{-p}\right)\left(\frac{a\nu}{2\pi}\right)^{1/2}e^{-a\nu/2}
\label{eq:shethor}
\end{equation}
with $\bar{\rho}_{\rm m} \equiv \Omega_{\rm m}\bar{\rho}$ the averaged
matter density of the Universe.
The shape of this mass function is parameterised by $a$ and $p$, and
$A$ is simply a normalisation factor.

\subsection{Halo correlation functions}
At sufficiently large separations,
the two-point correlation of halos of mass $m$ can be related
to that of the matter density through the bias function
\begin{equation}
\xi_2^{\rm h}(x_1,x_2,z)=b(m,z)^2  \xi_2(x_1,x_2,z), 
\end{equation}
where $\xi_2$ is the linear dark matter density autocorrelation at redshift of
interest.
The function $b(m,z)$ can be estimated analytically using
the Press-Schechter formalism \citep{MoWhite}. 
Here, to remain consistent with eq.~(\ref{eq:shethor}), 
we use the expression for the bias of \citet{ST99}
\begin{equation}
b(m,z) = 1 + \frac{a\nu-1}{\delta_{\rm c}} + \frac{2p/\delta_{\rm
    c}}{(1+a\nu)^p}.
\label{eq:bias}
\end{equation}
This result is valid in the regime where the separation  $x=|x_2-x_1|$ 
is large enough compared to the mass scale
$R(m)$. This has been tested successfully against $N$-body
simulations by \citet{MoWhite2} \citep[see however e.g.][for possible
improvements on eq.~\ref{eq:bias}]{Tinker&al2010}.

We obtain the bias of halos exceeding mass threshold $m$ by
calculating the weighted average
\begin{equation}
b(>m,z) = \frac{\int_m^\infty  b(m',z)n(m',z) {\rm d}m'}{\int_m^\infty n(m',z) {\rm d}m'},
\end{equation}
and hence the averaged two-point correlation function for halos
above the threshold,
\begin{equation}
\bar{\xi}_2^{\rm h}(>m,z) = b(>m,z)^2 {\bar \xi}_2(z).
\end{equation}
Recall that this equation should be valid in the regime where the patch size is
large compared to $R(m)$,
\begin{equation}
L \gg R(m)=\left(\frac{3\pi m}{4 {{\bar \rho}_{\rm m}}} \right)^{1/3},
\label{eq:condition}
\end{equation}
but small enough that light cone effects on the clustering inside it are negligible. 

\subsection{Generalised Extreme Value parameterisation}
\label{sec:GEV}
The method outlined above allows us to compute the complete distribution function of
of the most massive clusters. However, due to its complexity and the necessity of computing some
of the integrals numerically, it does not provide us with a neat analytic
parameterisation of the distribution. Therefore, in order to achieve
such a parameterisation, we turn to the general theory
of extremes, eq.~(\ref{eq:GEV}), using $u=\log_{\rm 10} m$ as
the random variable.  In order to calculate the parameters $\gamma$, $\alpha$
and $\beta$, we perform Taylor expansions of the analytic $P_{\rm GEV}$, and $P_0$ as computed by our
method in the Poisson regime (eq.~\ref{eq:p0poisson}).
This Taylor expansion is performed about the peaks
of the two distributions ${\rm d}P_0/{\rm d}u$ and ${\rm d}P_{\rm
  GEV}/{\rm d}u$. Equating the
first two terms in these expansions give us expressions for the three parameters:
\begin{eqnarray}
\gamma = n(>m_0)V-1, \quad \beta =
\frac{(1+\gamma)^{(1+\gamma)}}{n(m_0)Vm_0\ln 10}, \nonumber \\
\alpha = \log_{10} m_0 - \frac{\beta}{\gamma}[(1+\gamma)^{-\gamma} -1],
\end{eqnarray}
where $m_0$ is the mass at which the distribution ${\rm d}P_0/{\rm
  d}z=\ln(10)\ m\ p_{\rm G}(m)$ peaks -- hence close to the most likely
value of $m$ -- and is given implicitly by
\begin{eqnarray}
A \frac{\bar{\rho}_m V}{m_0} \sqrt{\frac{a}{2\pi \nu_0}}
e^{-a\nu_0/2} \left( 1+(a\nu_0)^{-p} \right)  =
\nonumber \\
 \frac{a}{2} + \frac{1}{2\nu_0} + \frac{ap
 (a\nu_0)^{-(p+1)}}{1+(a\nu_0)^{-p}} - \frac{\nu_0''}{\nu_0'^2} ,
\end{eqnarray}
where $\nu_0$, $\nu_0'$ and $\nu_0''$ are $\nu$ and its derivatives with respect to $m$
evaluated at $m=m_0$. 

These equations, then, allow us to neatly summarise the information contained in the
extreme value distribution with the single parameter $\gamma$ which describes its shape.
This statistic has the potential to be used as a tool to compare models with data or with
each other, as \cite{Mikelsons&al} proposed for the CMB.

\section{Numerical experiment}
\label{sec:HOR}
To test our halo mass Gumbel distribution we compare the analytical result
to measurements on the Horizon 4$\Pi$ Simulation \citep{Teyssier&al}, a large
cosmological dark matter simulation performed using the RAMSES $N$-body code
\citep{Teyssier}. The simulation followed the evolution of a cubic piece of the universe $2h^{-1}$Gpc
on a side containing $4096^3$ particles, i.e. with a  particle mass of $7.7h^{-1}\times 10^9 M_\odot$. 
Initial conditions were based on the WMAP 3-year results \citep{WMAP3},
with the Hubble constant, density and characteristic parameters
of the power spectrum given by ($h$, $\Omega_\Lambda$, $\Omega_{\rm m}$,
$\Omega_{\rm b}$, $\sigma_8$, $n_s$) = (0.73, 0.76, 0.24, 0.042, 0.77, 0.958).
Halos in the simulation were identified at present time, $z=0$, using a `Friends-of-Friends'
algorithm \citep[e.g.][]{Zel,Davis} with a standard linking length parameter value given by 0.2 times the
mean interparticle distance. 

\begin{figure}
  \centering
  \resizebox{\columnwidth}{!}{\includegraphics*[angle=-90]{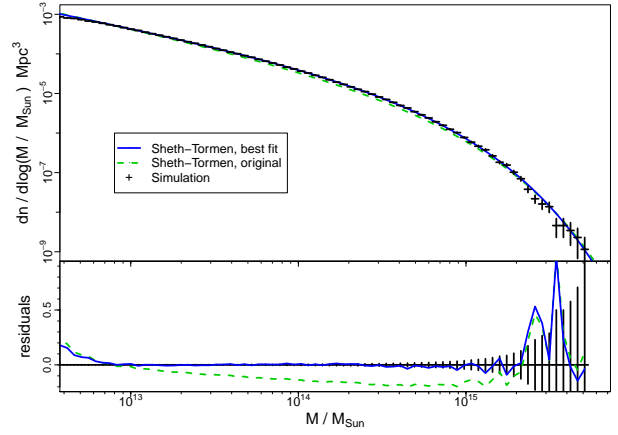}}
  \caption{\label{fig:massfunc}The upper panel shows the mass function of halos in the simulation
(points), compared to the Sheth-Tormen mass function with $(p,a)$ equal to 
$(0.3,0.707)$ and $(-0.19,0.777)$ (solid blue and dashed green lines
respectively). The lower panel shows the residuals of the two theoretical
curves compared to the data; i.e. (theory -- data)/data.}
\end{figure}

\subsection{Fit of the mass function}
Any discrepancies between our derived Gumbel distribution and the true
distribution can be thought of as arising from one of two sources:
either inaccuracies in our chosen mass function, or inaccuracies due
to the various assumptions made in proceeding from the mass function
to $p_{\rm G}$. In order to quantify the respective contributions of
each of these two sources, we repeat our calculations with two sets of parameters for
the mass function: (i) once taking the parameters used in
\citet{ST99}, $(p,a)=(0.3, 0.707)$, and, since 
the  Sheth \& Tormen form is its standard parametrisation
is known to perform only approximately \citep[e.g.][]{Warren,Jenkins&al}, 
(ii) once with a best fit for $a$ and $p$ to the simulation's mass function, leading
to $(p,a)=(-0.19,0.777)$. 
For this latter, we also weight bins by their mass, since
it is the high-mass end of the distribution which is of interest to us.
Fig. \ref{fig:massfunc} shows both these mass functions along with 
that measured in the simulation.

\begin{figure*}
  \centering
  \subfigure[$L = 100/h$ Mpc; $\gamma = -0.11$]{\label{fig:gumbel100}\resizebox{0.48\textwidth}{!}{\includegraphics*[angle=-90]{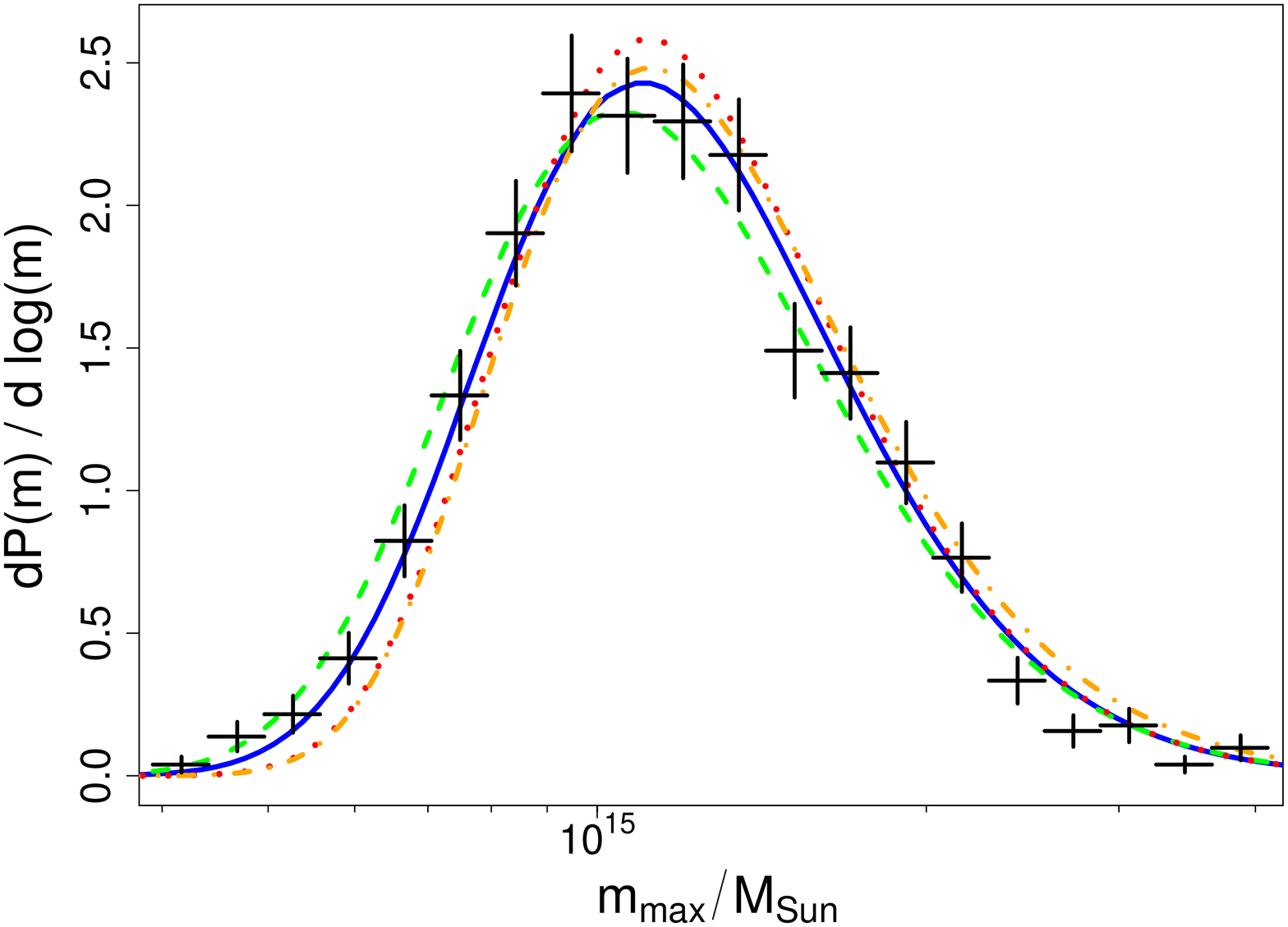}}}
  \subfigure[$L = 50/h$ Mpc; $\gamma = -0.14$]{\label{fig:gumbel50}\resizebox{0.48\textwidth}{!}{\includegraphics*[angle=-90]{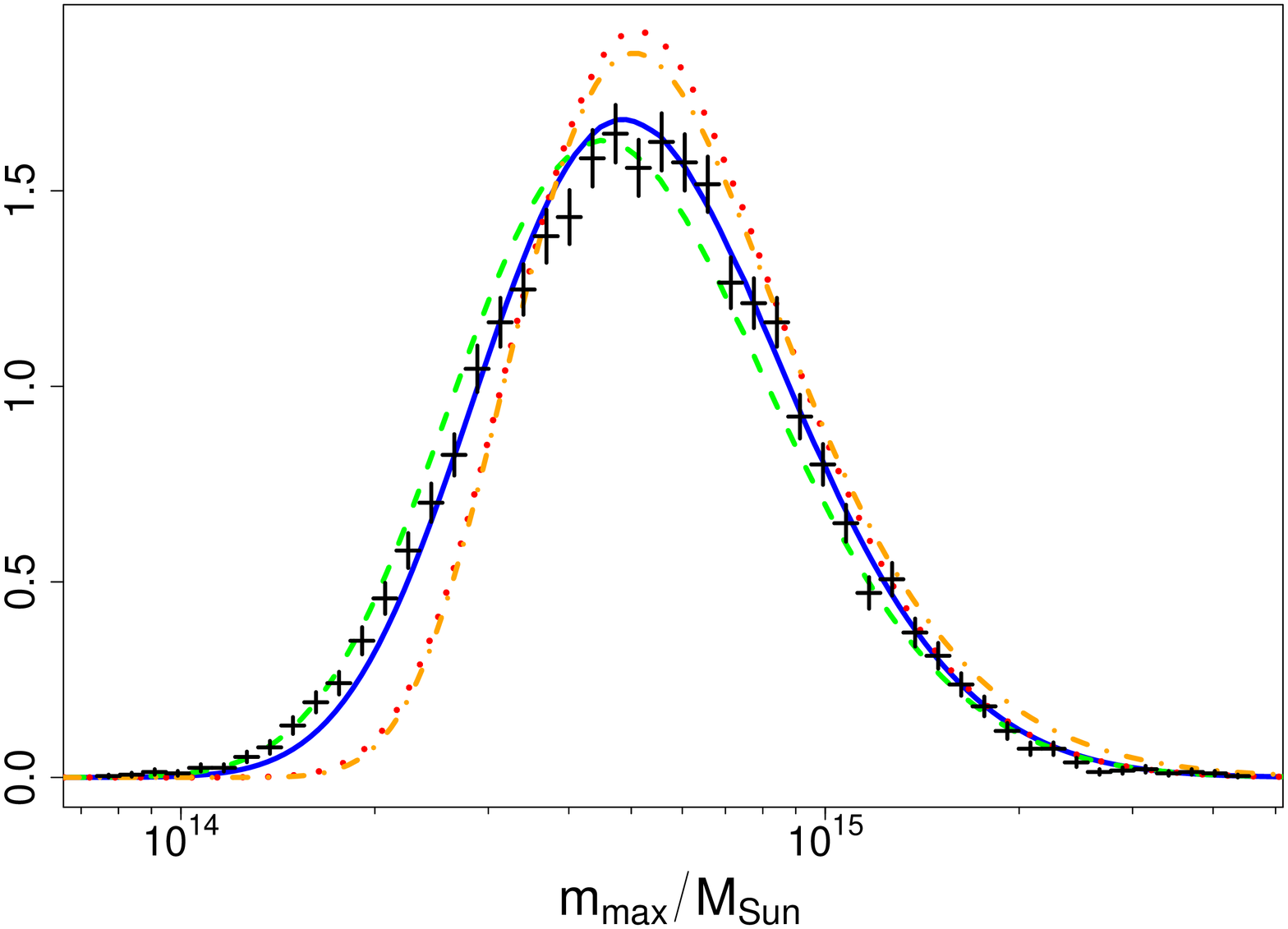}}}
  \subfigure[$L = 20/h$ Mpc; $\gamma = -0.11$]{\label{fig:gumbel20}\resizebox{0.48\textwidth}{!}{\includegraphics*[angle=-90]{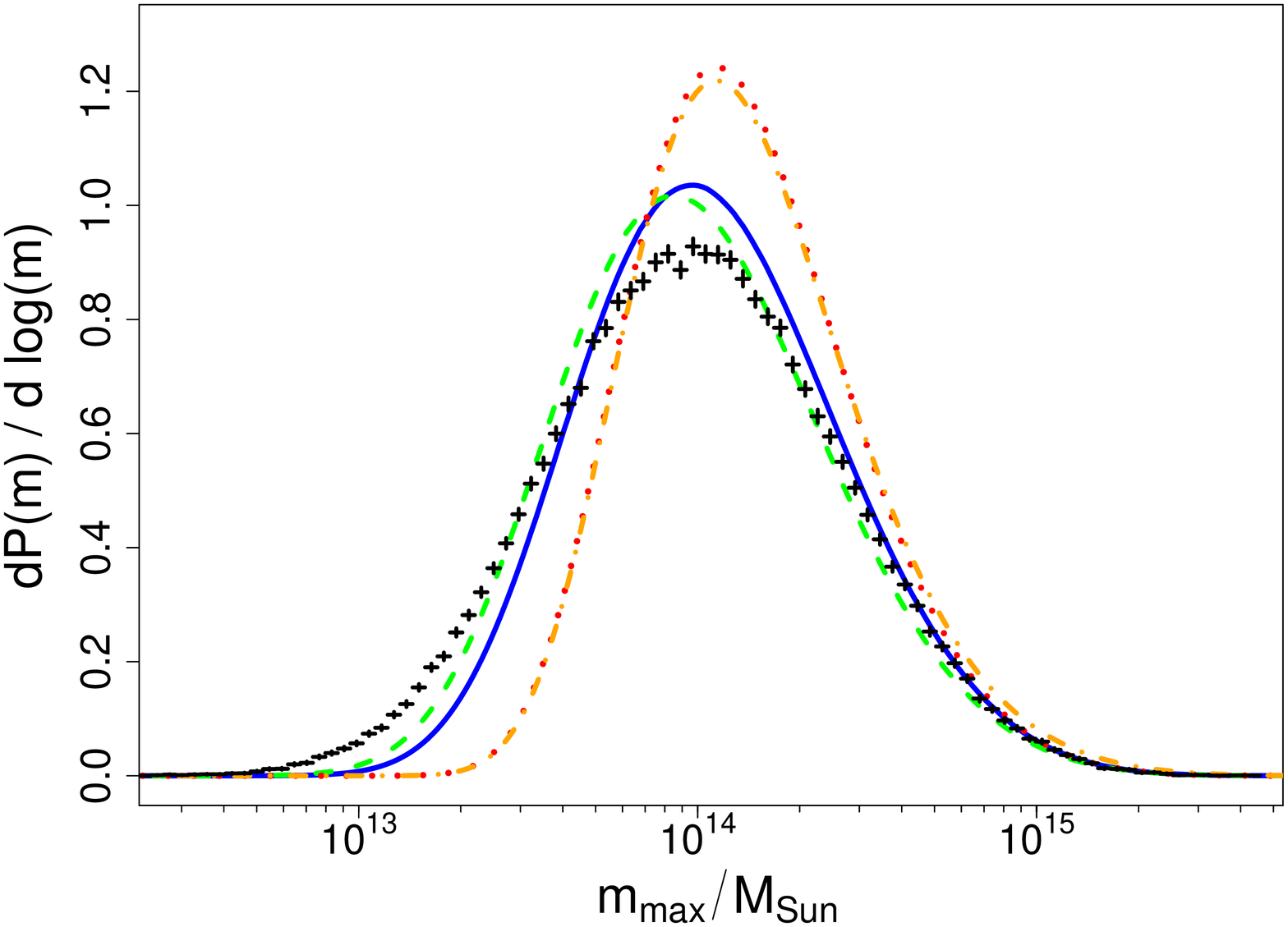}}}
  \subfigure[$L = 8/h$ Mpc; $\gamma = -0.12$]{\label{fig:gumbel8}\resizebox{0.48\textwidth}{!}{\includegraphics*[angle=-90]{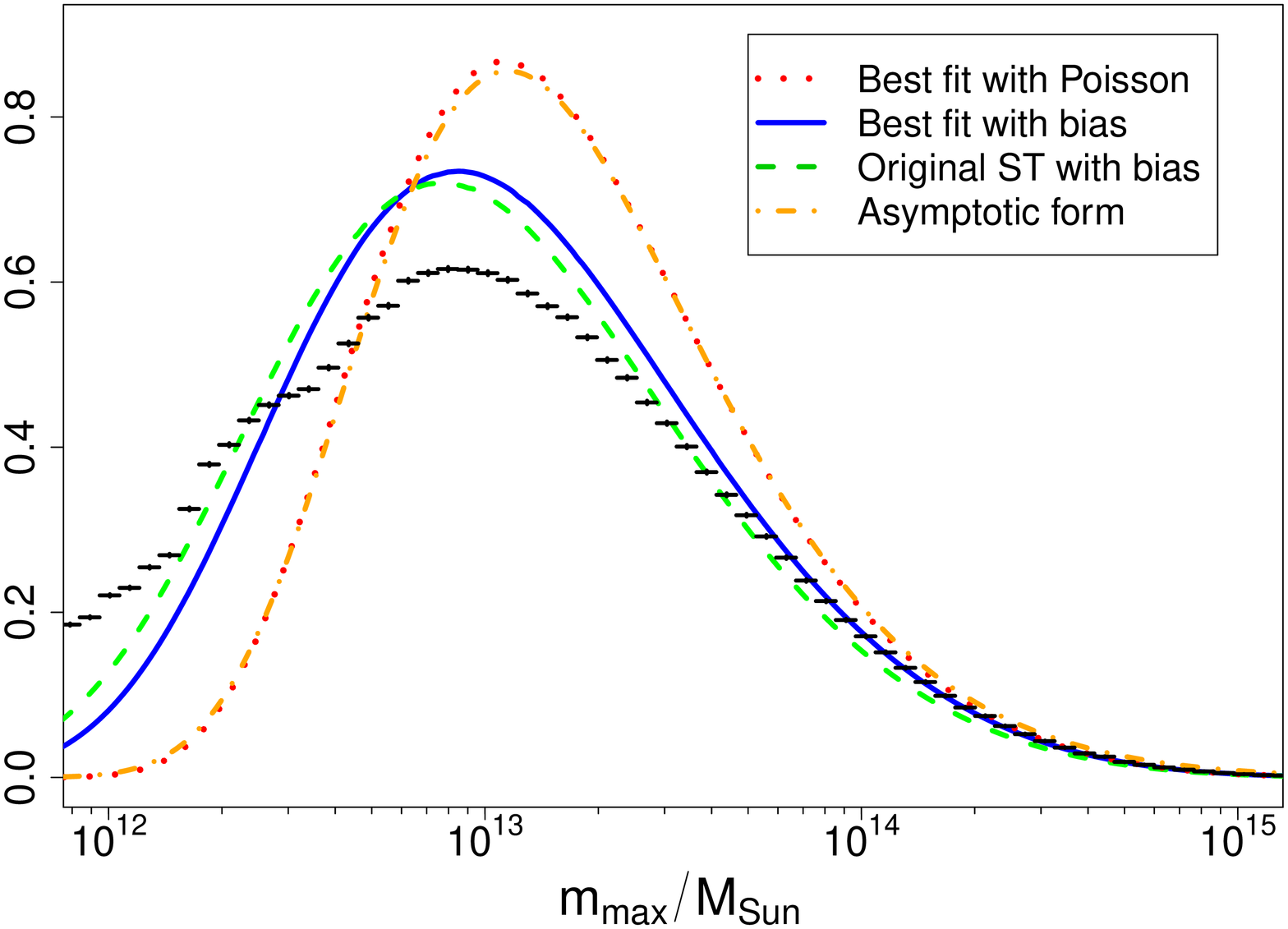}}}
  \caption{Distribution of largest cluster masses $m_{\rm max}$. The
    y axis is probability density per
log mass. Points are measurements from the simulation with Poisson error bars for
each mass bin. The solid blue line is the theoretical result using the best-fit mass function
(\mbox{$p=-0.19, a=0.777$}) and full halo clustering. Green dashes have instead the original Sheth-Tormen
parameters \mbox{($p=0.3, a=0.707$)}. Not surprisingly, they do not
agree as well with the measurement as the solid blue line; red dots have the best-fit values but assume halos are Poisson
distributed. The orange dot-dashed line is the Generalised Extreme Value distribution, with parameters
calculated as explained in section \ref{sec:GEV} and assuming Poisson statistics. All calculations
use spherical patches, with radius \mbox{$L=100,50,20,8 h^{-1}$ Mpc respectively.}}
  \label{fig:gumbel}
\end{figure*}

\subsection{Results}

Fig. \ref{fig:gumbel} shows the distribution $p_{\rm G}(m_{\rm max})$ calculated as above,
both for Poisson statistics (equation \ref{eq:p0poisson}, red dots) and
incorporating full clustering (equation \ref{eq:p0clustered}, blue solid lines) using our best-fit
mass function. The full clustering calculation  is also shown for the original Sheth-Tormen
parameters (green dashes).
Points show measurements from the Horizon 4$\Pi$ simulation for comparison.

Fitting a Gumbel distribution eq.~(\ref{eq:GEV}) to the data presented in the four 
panels of this figure yields a single value of $\gamma$ around -0.21 $\pm^{0.02}_{0.01}$
with reduced $\chi^2 \le 1.1$, whereas the analytic prediction presented in 
section \ref{sec:GEV} gives $ -0.14 \le \gamma \le -0.1$. This lack of agreement
has its root in the fact that $\gamma$ is very sensitive to the higher order (skewness 
and kurtosis) moments of the data distribution around its peak, and these are poorly 
captured by assuming Poisson statistics, even on $100h^{-1}$ Mpc scales.

However, for a patch (in this case a sphere of radius $L$) of size $L=100h^{-1}$ Mpc,
Fig. \ref{fig:gumbel100} shows that  $p_{\rm G}(m_{\rm max})$ measured in the data 
is not as badly described by Poisson theoretical results as it would have seemed from the value of 
$\gamma$ alone: we are closing in on the so-called ``scale of homogeneity'' above which the matter distribution is
essentially unclustered. Reducing the patch size to $50h^{-1}$ Mpc (fig. \ref{fig:gumbel50})
causes the full clustering and Poisson curves to diverge. As expected, only the calculations incorporating 
clustering remain a good match to the simulation on these smaller scales.

Decreasing $L$ further (fig. \ref{fig:gumbel20}, \ref{fig:gumbel8}) causes
even the calculations including clustering to diverge from the data as our approximations --
in particular the expression for the function $\sigma(y)$ in section
\ref{sec:gumbel} and the condition (\ref{eq:condition}) -- fail outside
the large-$L$ limit.  For instance, we find $R(10^{13} M_{\odot})=3$ Mpc/$h$,
and $R(10^{14} M_{\odot})=6.4$ Mpc/$h$ which is a significant fraction
of the respective patch sizes of $8$ Mpc/$h$ and $20$ Mpc/$h$. 
Despite these limitations, the description of the data is still significantly better
than that of the Poisson calculation and remains impressive at the
high-mass end. This is excellent news as it is this high-mass tail of
the distribution which is of interest for assessing the significance of rare events such as surprisingly massive clusters
observed in X-ray or redshift surveys. Indeed the lower-mass tail of
the distribution for which the prediction fails most significantly
lies at masses below $\sim10^{13}M_\odot$, which corresponds to halos
containing one to a few galaxies rather than tens or hundreds, and
therefore are of limited interest in the search for the most massive cluster.

\begin{figure}
  \centering
  \resizebox{\columnwidth}{!}{\includegraphics*[angle=-90]{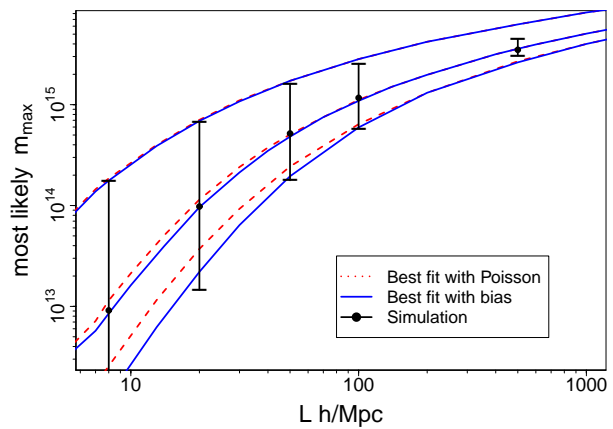}}
  \caption{The most likely value of $\log_{10} m_{\rm max}$ (middle line and points)
and the $95\%$ confidence limits (upper and lower line and bars).
Dashed red and solid blue lines are analytic results for the Poisson
limit and  full clustering calculations respectively, 
and points are simulation values for $L$ corresponding
to the four panels of fig. \ref{fig:gumbel} plus \mbox{$L=500/h$} Mpc.}
  \label{fig:max}
\end{figure}

Moreover, note that the position of the peak of the probability
distribution function - that is, the most likely value of
$\log_{10}m_{\rm max}$ - is fairly accurately predicted by the theory even when the shape
of the curve begins to diverge from that of the data. Fig. \ref{fig:max} shows this most likely value,
$\log_{10} \hat{m}_{\rm max}$, as a function of $L$ for both
theoretical estimates and the simulation data
(central line and points in the figure). We also show in this figure the $95\%$ confidence region on
$\log_{10} m_{\rm max}$ (upper and lower lines and bars). This too is well fit by the theory,
particularly for the upper limit, and we emphasize that this is a crucial test of the theory's
ability to give significance values for observations of specific
overly massive clusters.

In addition to the four values of $L$ shown in Fig. \ref{fig:gumbel}, Fig. \ref{fig:max}
has a final simulation point at \mbox{$L=500/h$ Mpc}. Here the $95\%$ confidence region
is poorly fit, because the patch is too large compared to the simulation box to
get good statistics. However, the measured value of $\log_{10} m_{\rm max}$ is still in good agreement
with the theory.

\subsubsection{Senstivity to the mass function}
Worthy of note is the close similiarity of the curves for the two sets of parameters in the Sheth-Tormen
mass function (solid blue and dashed green lines in Fig. \ref{fig:max}). Although the mass functions
themselves differ by a relatively large amount compared to the best-fit function's agreement with
the simulation points (Fig. \ref{fig:massfunc}, lower panel), this translates to only a modest change
in the distribution of $m_{\rm max}$. Similar calculations performed with the mass functions of
\citet{Tinker&al2008} and \citet{Jenkins&al} lead us to conclude that the extreme value distribution
is fairly robust to the choice of any reasonable analytic fit to the mass function.

\subsubsection{Redshift variation}
While the above calculations use patch sizes $L$ small enough that the redshift evolution within
the patch is negligible, it is also interesting to calculate $m_{\rm max}$ for a larger region
with significant $\Delta z$. As noted in section \ref{sec:gumbel}, redshift variation can be taken
into account by a weighted spatial average provided the Poisson approximation holds, which is the
case for such large patches. In particular, averaging the number density of halos over the range
$z=0$ to $\infty$ gives a value for the expected largest mass cluster in the entire observable
universe.

We performed this calculation assuming Poisson statistics, and found $m_{\rm max} = 4.6 \pm^{1.2}_{0.6} \times 10^{15} M_\odot$
at the $1-\sigma$ confidence level. We note that this is in fair agreement with the similar calculation
performed by \cite{HolzPerlmutter}, who obtained $m_{\rm max} = 3.8 \pm^{0.6}_{0.5}\times 10^{15}$
using a WMAP7 cosmology and the mass function of \cite{Tinker&al2008}.

\section{Conclusion and Discussion}
\label{sec:conclusion}
We have presented an analytic prediction of the the probability distribution
of $m_{\rm max}$, the most massive dark matter halo/galaxy cluster in a specified region of
the universe, making use of the counts-in-cells formalism. Our
calculation, valid for Gaussian initial conditions, is numerically consistent 
with that proposed by \cite{HolzPerlmutter} when performed assuming such massive halos 
are Poisson distributed spatially. 
However, the work presented in this paper improves on the calculation performed by these authors in
two aspects: (i) our results are given in a fully explicit analytic
form and (ii) they include the contribution of clustering of halos.

We also compared our analytic predictions to
measurements from a large (2000$h^{-1}$ Mpc on a side) and well
resolved  (particle mass $7.7h^{-1}\times 10^9$ M$_\odot$) cosmological $N$-body simulation
at zero redshift.
We achieve remarkable agreement with the simulation in the area of parameter
space in which our formalism is expected to be valid, namely patch radii above
a few tens of $h^{-1}$ Mpc. More surprisingly, even outside this range
of scales the high-mass tail of the distribution is well fit by our "fully clustered" theoretical estimate,
as is the most likely value of $\log_{10} m_{\rm max}$.

This unexpected success over a wide range of scales warrants the
application of the formalism to quantify the statistical significance
of individual clusters observed in surveys.  By applying our method to a patch of shape, size and redshift equivalent to
a real survey we can obtain the Gumbel distribution and hence a likelihood for
the observed value of $m_{\rm max}$. Moreover, we are quite
confident that our method can be extended to
non-standard cosmologies such as those including initial
non-Gaussianities. It could therefore provide a measure of the evidence
for such cosmologies from existing surveys of the most massive
clusters as advocated in \citet{Cayon&al10}. We plan to tackle this
exciting prospect in the very near future.

In addition to the full likelihood curve of $m_{\rm max}$ we are able, via the analytic
Generalised Extreme Value formalism, to produce a summary of the distribution in the form
of the three parameters $a$, $b$ and $\gamma$. The latter in particular has been proposed
previously as a statistic for use in model comparison \citep[e.g.][]{Mikelsons&al}.
Although the work described in this paper uses a single cosmology and power spectrum
and produces a roughly constant value of $\gamma$, the results of \citet{Colombi}
suggest that $\gamma$ should be quite sensitive to the shape of the power spectrum.
If the effect on $\gamma$ of the clustering can be fully quantified, we therefore
expect to be able to use it as a statistic for direct comparison of models with observation.
Likewise $a$, which is closely related to the peak $m_0$ of the distribution, may
prove a useful statistic since we have demonstrated that it is well predicted by
our theory.

\section{Acknowledgements}
The authors are grateful to an anonymous reviewer whose comments led to improvements in this paper.

JD's research is supported by the Oxford Martin School, Adrian Beecroft and STFC.

CP acknowledges support from a Leverhulme visiting professorship at the Astrophysics department of the University of Oxford.

\bibliographystyle{plain}
\bibliography{paper}
\bsp
\label{lastpage}
\end{document}